\documentstyle[12pt]{article}
\def\Re{{\cal R \mskip-4mu \lower.1ex \hbox{\it e}\,}}
\def\Im{{\cal I \mskip-5mu \lower.1ex \hbox{\it m}\,}}

\def\etal{{\it et al.}}

\def\beq{\begin{equation}}
\def\eeq{\end{equation}}
\def\brbsg{B(b\to s\gamma)}
\def\bsg{b\to s\gamma}
\def\Zbb{Z\rightarrow b\ov b}
\def\epsb{\epsilon_b}
\def\eps1{\epsilon_1}
\def\eps2{\epsilon_2}
\def\eps3{\epsilon_3}
\def\sub#1{_{\lower.25ex\hbox{$\scriptstyle#1$}}}
\def\sul#1{_{\kern-.1em#1}}
\def\sll#1{_{\kern-.2em#1}}
\def\sbl#1{_{\kern-.1em\lower.25ex\hbox{$\scriptstyle#1$}}}
\def\ssb#1{_{\lower.25ex\hbox{$\scriptscriptstyle#1$}}}
\def\sbb#1{_{\lower.4ex\hbox{$\scriptstyle#1$}}}

\def\GeV{\,{\rm GeV}}

\def\JL{J. L. Lopez}
\def\DVN{D. V. Nanopoulos}

\def\to{\rightarrow}
\def\ov{\overline}
\def\mh{\ifmmode m\sbl H \else $m\sbl H$\fi}
\def\mch{\ifmmode m_{H^\pm} \else $m_{H^\pm}$\fi}
\def\mt{\ifmmode m_t\else $m_t$\fi}
\def\mc{\ifmmode m_c\else $m_c$\fi}
\def\mz{\ifmmode M_Z\else $M_Z$\fi}
\def\mw{\ifmmode M_W\else $M_W$\fi}
\def\mws{\ifmmode M_W^2 \else $M_W^2$\fi}
\def\mhs{\ifmmode m_H^2 \else $m_H^2$\fi}
\def\mzs{\ifmmode M_Z^2 \else $M_Z^2$\fi}
\def\mts{\ifmmode m_t^2 \else $m_t^2$\fi}
\def\mcs{\ifmmode m_c^2 \else $m_c^2$\fi}
\def\mchs{\ifmmode m_{H^\pm}^2 \else $m_{H^\pm}^2$\fi}
\def\ztwo{\ifmmode Z_2\else $Z_2$\fi}
\def\zone{\ifmmode Z_1\else $Z_1$\fi}
\def\mtwo{\ifmmode M_2\else $M_2$\fi}
\def\mone{\ifmmode M_1\else $M_1$\fi}
\def\tb{\ifmmode \tan\beta \else $\tan\beta$\fi}
\def\xw{\ifmmode x\sub w\else $x\sub w$\fi}
\def\ch{\ifmmode H^\pm \else $H^\pm$\fi}
\def\lum{\ifmmode {\cal L}\else ${\cal L}$\fi}
\def\inpb{\ifmmode {\rm pb}^{-1}\else ${\rm pb}^{-1}$\fi}
\def\infb{\ifmmode {\rm fb}^{-1}\else ${\rm fb}^{-1}$\fi}
\def\epem{\ifmmode e^+e^-\else $e^+e^-$\fi}
\def\ppb{\ifmmode \bar pp\else $\bar pp$\fi}

\newskip\zatskip \zatskip=0pt plus0pt minus0pt

\def\matth{\mathsurround=0pt}
\def\lsim{\mathrel{\mathpalette\atversim<}}
\def\gsim{\mathrel{\mathpalette\atversim>}}
\def\atversim#1#2{\lower0.7ex\vbox{\baselineskip\zatskip\lineskip\zatskip
  \lineskiplimit 0pt\ialign{$\matth#1\hfil##\hfil$\crcr#2\crcr\sim\crcr}}}

\renewcommand{\thefootnote}{\fnsymbol{footnote}}

\begin{document} \begin{titlepage}
\setcounter{page}{1}
\thispagestyle{empty}
\rightline{\vbox{\halign{&#\hfil\cr
&YUMS-95-9\cr
&SNUTP-95-030\cr
&hep-ph/9504369\cr
&March 1995\cr}}}
\vspace{0.1in}
\begin{center}
\vglue 0.3cm
{\Large\bf Implications of the Recent Top Quark Discovery\\}
\vspace{0.2cm}
{\Large\bf on Two Higgs Doublet Model\\}
\vglue 1.5cm
{GYE T. PARK\\}
\vglue 0.4cm
{\em Department of Physics, Yonsei University\\}
{\em Seoul, 120-749, Korea\\}
\baselineskip=12pt

\end{center}

\begin{abstract}

Concentrating on the impact of the very recent top quark discovery, we perform
a combined analysis of two strongest constraints
on the 2 Higgs doublet model, one coming from the recent measurement by CLEO
on the inclusive branching ratio of $b\rightarrow s\gamma$ decay and the other
from the recent LEP data on $Z\rightarrow b\overline b$ decay.
We have included the model predictions for one-loop vertex corrections to
$\Zbb$ through $\epsb$.
We find that the $\epsb$ constraint excludes most of the less appealing window
$\tan\beta\lsim 1$ at $95\%$C.~L. for the measured top mass from CDF,
$m_t=176\pm 8\pm10\;{\rm GeV}$.
Moreover, it excludes  $\tan\beta\lsim 2$ at $95\%$C.~L. for $m_t\gsim176\;{\rm
GeV}$.
Combining with the $\bsg$ constraint, only very heavy charged Higgs ($\gsim
670\;{\rm GeV}$) is allowed by the measured $m_t$ from CDF.

\end{abstract}

\renewcommand{\thefootnote}{\arabic{footnote}} \end{titlepage}
\setcounter{page}{1}


Very recently, the CDF Collaboration from Fermi Laboratory has finally
announced their observation of top quark production in $\ov{p} p$ collisions
with the measured top mass \cite{CDF-topdiscovery}, $m_t=176\pm 8\pm10\;{\rm
GeV}$. The top quark discovery now leaves the Standard Higgs mass $m_H$
the only unknown parameter in the Standard Model(SM).
The unknown $m_t$ has long been one of the biggest disadvantages in studying
the phenomenology of the SM and its extensions of interest.
Now that $m_t$ becomes known at last, one should be able to narrow down
the values of $m_t$ in the vicinity of the above central value.
Despite the remarkable successes of the SM in its complete
agreement with current all experimental data, there is still no
experimental information on the nature of its Higgs sector.
The 2 Higgs doublet model(2HDM) is one of the mildest extensions of the SM,
which has been consistent with experimental data.
In this letter, we would like to present the implications of the top quark
discovery on the 2HDM in view of the two strongest constraints present in the
model, namely, the ones from the flavor-changing radiative decay $b\rightarrow
s\gamma$ and $Z\rightarrow b\overline b$ decay.
In the 2HDM to be considered here, the Higgs sector consists of 2 doublets,
$\phi_1$ and $\phi_2$,
 coupled to the charge -1/3 and +2/3 quarks, respectively, which will ensure
the absence of Flavor-Changing Yukawa couplings at the tree level
 \cite{NOFC}. The physical Higgs spectrum of the model includes two CP-even
 neutral Higgs($H^0$, $h^0$), one CP-odd neutral Higgs($A^0$)
, and a pair of charged Higgs($H^\pm$). In addition to the masses of these
Higgs, there is another free parameter in the model, which is $\tan\beta\equiv
v_2/v_1$, the ratio of the vacuum expectation values of both doublets.

After the first observation by CLEO on the exclusive decay $B\rightarrow
K^*\gamma$\cite{Thorndike}, CLEO has recently measured for the first time
the inclusive branching ratio of $b\rightarrow s\gamma$ decay
to be at 95\% C.~L. \cite{CLEO94},
$$1\times 10^{-4}<B(b\rightarrow s\gamma) <4\times 10^{-4}.$$
This follows the renewed surge of interests on the $\bsg$ decay, spurred by the
CLEO bound $\brbsg<8.4\times10^{-4}$ at $90\%$ C.L. \cite{CLEO}, with which
it was pointed out in Ref.~ \cite{BargerH} that the CLEO bound can be violated
due to the charged Higgs
contribution in the 2HDM and the Minimal Supersymmetric Standard Model(MSSM)
basically if $m_{H^\pm}$ is too light, excluding large portion of the charged
Higgs parameter space. It has certainly proven that this particular decay mode
can provide more stringent constraint on new physics beyond SM than any other
experiments\cite{bsgamma}. However, it turns out in the 2HDM that the only
constraint competing
with the one from $b\rightarrow s\gamma$ comes from the LEP data on the
$Z\rightarrow b\overline b$ decay\cite{Rbbsg2HD}.
As we know, with the increasing accuracy of the LEP measurements, it has become
extremely important performing the precision test of the SM and its
extensions\footnote{A standard model fit to the latest LEP data yields the top
mass,
$m_t=178\pm 11^{+18}_{-19}\;{\rm GeV}    $  \cite{Schaile}, which is in perfect
agreement with the measured top mass from CDF.}.
Among several different schemes to analyze precision electroweak tests, we
choose a scheme introduced by Altarelli et.~al.
\cite{ABC,Altlecture} where four variables, $\epsilon_{1,2,3}$ and $\epsilon_b$
are defined in a model independent way. These four variables correspond to
a set of observables $\Gamma_{l}, \Gamma_{b}, A^{l}_{FB}$ and $M_W/M_Z$.
Among these variables, $\epsilon_b$ encodes the vertex corrections to
$Z\rightarrow b\overline b$.

In the 2HDM and the MSSM, $b\rightarrow s\gamma    $ decay receives significant
contributions from penguin diagrams with $W^\pm-t$ loop, $H^\pm-t$ loop
\cite{HCbsg} and the $\chi^\pm_{1,2}-\tilde t_{1,2}$ loop \cite{Bertolini}
only in the MSSM.
The expression used for $B(b\rightarrow     s\gamma)    $ in the leading
logarithmic (LL) calculations is given by \cite{GSW}
\begin{equation}
{B(b\rightarrow     s\gamma)\over B(b\rightarrow
ce\bar\nu)}={6\alpha\over\pi}
{\left[\eta^{16/23}A_\gamma
+{8\over3}(\eta^{14/23}-\eta^{16/23})A_g+C\right]^2\over
I(m_c/m_b)\left[1-{2\over{3\pi}}\alpha_s(m_b)f(m_c/m_b)\right]},\label{bsg}
\end{equation}
where $\eta=\alpha_s(M_W)/\alpha_s(m_b)$, $I$ is the phase-space factor
$I(x)=1-8x^2+8x^6-x^8-24x^4\ln x$, and $f(m_c/m_b)=2.41$ the QCD
correction factor for the semileptonic decay.
$C$ represents the leading-order QCD
corrections to the $b\rightarrow s\gamma    $ amplitude when evaluated at the
$\mu=m_b$ scale
\cite{GSW}.
We use the 3-loop expressions for $\alpha_s$ and choose $\Lambda_{QCD}$ to
obtain $\alpha_s(M_Z)$ consistent with the recent measurements at LEP.
In our computations we have used: $\alpha_s(M_Z)=0.118$, $ B(b\rightarrow
ce\bar\nu)=10.7\%$, $m_b=4.8\;{\rm GeV}    $, and
$m_c/m_b=0.3$. The $A_\gamma,A_g$ are the
coefficients of the effective $bs\gamma$ and $bsg$ penguin operators
evaluated at the scale $M_W$. Their simplified expressions are given in
Ref.~\cite{BG}
in the justifiable limit of negligible gluino and neutralino contributions
\cite{Bertolini} and degenerate squarks, except for the $\tilde t_{1,2}$ which
are significantly split by $m_t$.
Regarding large uncertainties in the LL QCD corrections, which is mainly due to
the choice of renormalization scale $\mu$ and is estimated to be $\approx
25\%$, it has been recently demonstrated by Buras {\it et al.} in
Ref.~\cite{burasetal}
that the significant $\mu$ dependence in the LL result can in fact be reduced
considerably by including next-to-leading logarithmic (NLL) corrections, which
however, involves very complicated calculations of three-loop mixings
between cetain effective operators and therefore have not been completed yet.
In Fig. 1 we present the excluded regions in ($m_{H^\pm}$, $\tan\beta$)-plane
in the 2HDM
for $m_t=163$, and $176\GeV$, which lie to the left of each dotted curve.
The $m_t$ values are of course the central value and the lower limit from the
CDF. The contours are obtained using the new 95\% C.~L. upper bound
$B(b\rightarrow s\gamma) =4\times 10^{-4}$.
As expected, the new CLEO bound excludes a large portion of the parameter
space. We have also imposed in the figure the lower bound on $\tan\beta$ from
${m_t\over{600}}\lsim\tan\beta\lsim{600\over{m_b}}$ obtained by demanding that
the theory remain perturbative\cite{BargerLE}.
We see from the figure that at large $\tan\beta$ one can obtain a lower bound
on $m_{H^\pm}$ for each value of $m_t$. And we obtain the bounds
, $m_{H^\pm}\gsim 672, 843\GeV$ for $m_t=163, 176\GeV$, respectively.

Following Ref.~\cite{ABC}, $\epsilon_b    $ is defined from $\Gamma_b$, the
inclusive
partial width for $Z\rightarrow b\overline b    $, as
\begin{equation}
\epsilon_b    ={g^b_A\over{g^l_A}}-1
\end{equation}
where $g^b_A$ $(g^l_A)$ is the axial-vector coupling of $Z$ to $b$ $(l)$.
In the SM, the diagrams for $\epsilon_b    $  involve top quarks and
$W^\pm$ bosons  \cite{RbSM}, and the contribution to $\epsilon_b    $ depends
quadratically on $m_t$ ($\epsilon_b    =-G_F m_t^2/4\sqrt {2}\pi^2 + \cdots$).
In supersymmetric models there are additional diagrams
involving Higgs bosons and supersymmetric particles. The charged Higgs
contributions have been calculated in Refs.~ \cite{Denner,Rbbsg2HD} in
the context of the 2HDM, and the
contributions involving supersymmetric particles in Refs.~ \cite{BF,Rb2HD}.
The main features of the additional supersymmetric contributions are: (i) a
negative contribution
from charged Higgs--top exchange which grows as $m^2_t/\tan^{2}\beta$ for
$\tan\beta\ll{m_t\over{m_b}}$; (ii) a positive contribution from chargino-stop
exchange which in this case grows as $m^2_t/\sin^{2}\beta$; and (iii) a
contribution from neutralino(neutral Higgs)--bottom exchange which grows as
$m^2_b\tan^{2}\beta$ and is negligible except for large values of $\tan\beta$
({\it i.e.}    , $\tan\beta\gsim{m_t\over{m_b}}$).
$\epsb$ is closely related to the real part of the vertex correction to $\Zbb$
, $\nabla_b$
defined in Ref\cite{BF}.
The additional
diagrams involving $H^\pm$ bosons have been calculated in
Ref\cite{Rbbsg2HD,Rb2HD,BF,Denner}. The charged Higgs contribution to
$\nabla_b$ is given as \cite{BF}
\begin{equation}
\nabla_b^{H^\pm}={\alpha\over 4\pi \sin^2\theta_W}\left[
{2 v_L F_L+2 v_R F_R}\over {v_L^2+v_R^2}
\right] \;,
\end{equation}
where $F_{L,R}=F_{L,R}^{(a)}+F_{L,R}^{(b)}+F_{L,R}^{(c)}$ and
\begin{eqnarray}
F_{L,R}^{(a)} &=& b_1\left(M_{H^+}, m_t, m_b\right) v_{L,R}
\lambda^2_{L,R}\;,\\
F_{L,R}^{(b)} &=&\left[\left({M_Z^2\over{\mu^2}}
c_6\left(M_{H^+}, m_t, m_t\right)-{1\over 2}-c_0\left(M_{H^+}, m_t,
m_t\right)\right)v_{R,L}^t\right. \nonumber \\
&& \hspace*{1.05in} \left. +{m_t^2\over{\mu^2}}
c_2\left(M_{H^+}, m_t, m_t\right)v_{L,R}^t\right]\lambda^2_{L,R}\;,\\
F_{L,R}^{(c)} &=& c_0\left(m_t, M_{H^+}, M_{H^+}\right)\left({1\over 2}-
\sin^2\theta_W\right)\lambda^2_{L,R}\;,
\end{eqnarray}
where $\mu$ is the renormalization scale and
\begin{eqnarray}
v_L &=& -{1\over 2}+{1\over 3}\sin^2\theta_W\,,  \quad v_R={1\over
3}\sin^2\theta_W \;, \\
v_L^t &=& {1\over 2}-{2\over 3}\sin^2\theta_W\,,  \quad v_R^t=-{2\over
3}\sin^2\theta_W \;, \\
\lambda_L &=& {m_t\over{{\sqrt 2} M_W \tan\beta}}\,,
\quad\lambda_R = {m_b \tan\beta\over{{\sqrt 2} M_W }}\;.
\end{eqnarray}
The $b_1$ and $c_{0,2,6}$ above are the reduced Passarino-Veltman
functions\cite{BF,Ahn}.
In our calculation, we neglect  the neutral Higgs contributions to $\nabla_b $
which are all proportional to $m_b^2\tan^2\beta$ and
become sizable only for
$\tan\beta>{m_t\over{m_b}}$ and very light neutral Higgs $\lsim50\GeV$, but
decreases rapidly to get negligibly small as the Higgs masses become
$\gsim100\GeV$\cite{Denner}.
We also neglect oblique corrections from the Higgs bosons just to avoid
introducing more paramters. However, this correction can become sizable
when  there are large mass splittings between the charged and neutral Higgs,
for example, it can grow as $m^2_{H^\pm}$ if
$m_{H^\pm}\gg m_{H^0,h^0,A^0}$.
Although $\tan\beta\gg1$ seems more appealing because of apparent hierarchy
 $m_t\gg m_b$, there are still no convincing arguments against $\tan\beta<1$.
 Our goal here is to see if one can put a severe constraint from $\epsb$ in
this region.
In Fig. 1 we show  the contours (solid) of a predicted value of
$\epsb=-0.00733$,
which is the LEP lower limit at $95\%$C.~L.\cite{Altarelli94}.
The excluded regions lie below each solid curve for given $m_t$.
For $m_t=176(163)\GeV$, $\tan\beta\lsim 2.0(0.6)$ is ruled out at $95\%$C.~L.
for $m_{H^\pm}\lsim 1000\GeV$. We note that these strong constraints for
$\tan\beta\lsim 1$
stem from large deviations of $\epsb$ from the SM prediction, which grows as
$m^2_t/\tan^{2}\beta$ as explained above.
Combining both $\bsg$ and $\epsb$ constraints, only the region above the solid
curve and to the right of the dotted curve survive.
For $m_t=176(163)\GeV$, $\tan\beta\gsim 2.0(0.6)$ and $m_{H^\pm}\gsim
843(672)\GeV$ are allowed at $95\%$C.~L.

We have also considered other constraints from low-energy data primarily in
$B-\ov{B}, D-\ov{D}, K-\ov{K}$ mixing that exclude low values of
$\tan\beta$\cite{BargerLE,LowEdata}. But it turns out that none of them can
hardly compete with the present $\epsb$ constraint\cite{assume1}.
Nevertheless, the CLEO
 bound is still by far the strongest constraint present in the charged Higgs
sector of the model for $\tan\beta> 1$. Therefore, we find that $\bsg$ and
$\epsb$ serve as the presently strongest and complimentary constraints in 2HDM.

In conclusion, we study the implications of the top quark discovered very
recently by CDF by performing a combined analysis of two strongest constraints
on the 2 Higgs doublet model, one coming from the recent measurement by CLEO
on the inclusive branching ratio of $b\rightarrow s\gamma$ decay and the other
from the recent LEP data on $Z\rightarrow b\overline b$ decay.
We have included the model predictions for one-loop vertex corrections to
$\Zbb$ through $\epsb$.
We find that the $\epsb$ constraint excludes most of the less appealing window
$\tan\beta\lsim 1$ at $95\%$C.~L. for the measured top mass, $m_t=176\pm
8\pm10\;{\rm GeV}$.
Moreover, it excludes  $\tan\beta\lsim 2$ at $95\%$C.~L. for $m_t\gsim176\;{\rm
GeV}$.
Combining with the $\bsg$ constraint, only very heavy charged Higgs ($\gsim
670\;{\rm GeV}$) is allowed by the measured $m_t$ from CDF.

\vskip.25in
\centerline{ACKNOWLEDGEMENTS}

The author thanks Professor Jihn E.~Kim for very helpful discussions.
This work has been supported in part by Yonsei University Faculty Research
Grant and in part by the Center for Theoretical Physics, Seoul National
University.


%
\def\NPB#1#2#3{Nucl. Phys. B {\bf#1} (19#2) #3}
\def\PLB#1#2#3{Phys. Lett. B {\bf#1} (19#2) #3}
\def\PLIBID#1#2#3{B {\bf#1} (19#2) #3}
\def\PRD#1#2#3{Phys. Rev. D {\bf#1} (19#2) #3}
\def\PRL#1#2#3{Phys. Rev. Lett. {\bf#1} (19#2) #3}
\def\PRT#1#2#3{Phys. Rep. {\bf#1} (19#2) #3}
\def\MODA#1#2#3{Mod. Phys. Lett. A {\bf#1} (19#2) #3}
\def\IJMP#1#2#3{Int. J. Mod. Phys. A {\bf#1} (19#2) #3}
\def\TAMU#1{Texas A \& M University preprint CTP-TAMU-#1}
\def\ARAA#1#2#3{Ann. Rev. Astron. Astrophys. {\bf#1} (19#2) #3}
\def\ARNP#1#2#3{Ann. Rev. Nucl. Part. Sci. {\bf#1} (19#2) #3}

\newpage

%
{\bf Figure Captions}
\begin{itemize}

\item Figure 1: The regions in $(m_{H^\pm},\tan\beta)$ plane excluded in 2HDM
 by the new CLEO bound at $95\%$C.~L. $\brbsg<4.0\times10^{-4}$, for $m_t=163,
176 \GeV$. The excluded regions
lie to the left of each dotted curve. The excluded regions by the latest LEP
value at $95\%$C.~L
$\epsb=-0.00733$. lie below each solid curve.
The values of $m_t$ used are as indicated.
\end{itemize}


\begin{thebibliography}{99}
\bibitem{assume1} We have assumed that the updated analysis would not change
the conclusions in Ref\cite{BargerLE} drastically.
\end{thebibliography}
\end{document}